# Topological gap solitons in Rabi Su-Schrieffer-Heeger lattices

Chunyan Li,[1,2,*] Yaroslav V. Kartashov[2]

[1]*School of Physics, Xidian University, Xi'an, 710071, China*
[2]*Institute of Spectroscopy, Russian Academy of Sciences, 108840, Troitsk, Moscow, Russia*
*\*Corresponding author: chunyanli@xidian.edu.cn*

**Abstract**

Topological insulators are unique materials possessing forbidden topological gap and behaving similar to usual insulators in their bulk, but at the same time supporting localized in-gap states at their edges that demonstrate exceptional robustness, because they are protected by topology of the system and cannot be destroyed by local disorder or edge perturbations. Topological insulators discovered in different areas of physics, including optics, acoustics, and physics of Bose-Einstein and polariton condensates, usually employ potential landscapes (or lattices), designed such as to feature specific degeneracies in their linear modal spectra, that can be lifted by various physical effects or controllable deformations of the potential, leading to opening of the topological gap, where edge states appear. In this work, using binary Bose-Einstein condensate we propose a new type of topological insulator that does not explicitly use specially designed potential landscape, but instead utilizes spatially inhomogeneous Rabi coupling between two components, in the form of one- or two-dimensional Su–Schrieffer–Heeger (SSH) structure, combined with Zeeman splitting. Such Rabi lattices reveal the appearance of topologically nontrivial phases (including higher-order ones) controlled by spatial shift of the domains with enhanced coupling between condensates within unit cells of the structure, where localized topological states appear at the edges or in the corners of truncated Rabi lattice. We also show that the properties of edge states, their spatial localization, and location of their chemical potential within topological gap can be controlled by interatomic interactions that lead to formation of gap topological edge solitons bifurcating from linear edge states. Such solitons in condensates with inhomogeneous Rabi coupling appear as very robust nonlinear topological objects that do not require any threshold norm for their formation even in two-dimensional geometries, and that can exist in stable form for both attractive and repulsive interactions. Our results demonstrate considerable enhancement of stability of solitons in topological Rabi lattices in comparison with trivial Rabi lattices. They open new prospects for realization of topologically nontrivial phases by spatial engineering of coupling in multicomponent systems.

## Introduction

The phenomenon of topological insulators and closely associated with it formation of topologically protected edge states attracts nowadays enormous interest in diverse areas of science beyond solid-state physics, where such materials were initially discovered [1, 2]. Topological insulators were observed in mechanics [3], acoustics [4-6], on different linear [7-10] and nonlinear [11] photonic platforms, including photonic crystals [12-16] and waveguide arrays [17-21], in low-dimensional systems [22], in dissipative polariton condensates in microcavities [23-28], and in atomic systems in optical lattices [29-39]. Most of these realisations of topological insulators employ specially designed potentials, where spectra of linear eigenmodes possess degeneracies, such as Dirac points, that split under the action of certain physical effects (for example, in the presence of spin-orbit coupling and Zeeman splitting in atomic systems [40,41] or due to waveguide twisting in photonic systems [17]) or upon controllable deformation of the potential, preserving its periodicity (such as introduction of detuning of sublattices forming potential in valley-Hall systems [20]) leading to opening of the topological gap in linear spectrum, where edge states may appear if the insulator is truncated. Among the paradigmatic and simplest systems supporting topological edge states are the SSH-like structures [42], created recently also in atomic systems [33-35], and their extensions to two-dimensional geometries representing examples of higher-order topological insulators (see, e.g., works [43-47] and recent review [48]).

If the medium, where topologically nontrivial structure is created, is in addition nonlinear, the possibilities for control of evolution of topological edge states expand substantially, since nonlinear response allows to tune the location of edge states in the gap offering control over their localization degree [49], see also the review [11]. Strong confinement of topological edge states leads to the enhancement of parametric interactions involving such states, as was already observed experimentally [50-52]. In nonlinear medium, nontrivial topological phases can be induced by large-amplitude excitations [53-56], even if linear structure is topologically trivial. Most remarkably, nonlinear topological systems can also support a broad spectrum of topological edge solitons, usually bifurcating from linear edge states, as reported in various SSH-like atomic [57-59], mechanical [60], conservative [61-67] and dissipative [68-70] chains, in Floquet systems supporting unidirectional solitons [71-79], topological fibre loops [80], valley-Hall systems [81-83], photonic graphene [84], and higher-order insulators [85-87], to mention a few examples.

As mentioned above, usually topological systems rely on specially designed potentials, so the principal question arises whether the system with nontrivial topological properties can be realized without designing special potential landscape. In this work we propose such a system that is represented by a binary [88-89] Bose-Einstein condensate (BEC) with spatially inhomogeneous and topologically nontrivial Rabi coupling landscape between components. A linear interconversion (Rabi coupling) between the species, usually represented by two different spin atomic states, can be realized experimentally with existing techniques, for example, by applying a resonant electromagnetic field to condensate [90,91]. Remarkably, Rabi coupling can be made spatially inhomogeneous: When it is periodic, it creates so-called Rabi lattice, where nontopological one-dimensional gap solitons can form [92]. Rabi coupling can be also made periodic in time [93] imposing stabilizing action on multidimensional solitons even for attractive interactions.

Here we predict that when Rabi coupling landscape in binary BEC has the form of SSH lattice, whose unit cells feature two domains (spots) with enhanced coupling between species, one can realize topologically nontrivial phase by shifting these domains towards the edges of the unit cell. In this topological phase, localized states emerge at the edges of truncated Rabi lattice. We consider both one- (1D) and two-dimensional (2D) versions of Rabi lattices, the latter being the first example of higher-order topological insulator created by inhomogeneous Rabi coupling and supporting topological corner modes. Topological nature of these states is confirmed by calculation of topological invariants for bands of SSH Rabi lattice. Moreover, we show that in the presence of attractive or repulsive interactions this system supports topological edge solitons bifurcating in the gap from linear edge states. Such solitons show unexpected enhanced stability even in comparison with 1D gap solitons in nontopological Rabi lattices [92] and even in 2D settings, illustrating strong stabilizing role of inhomogeneous coupling, that is particularly important in multidimensional geometries [94-96].

**The Model**

The evolution of two components of binary BEC in the presence of spatially inhomogeneous Rabi coupling and Zeeman splitting can be described by the dimensionless Gross-Pitaevskii equation for the mean-field wavefunction $\mathbf{\Psi}(x,y,t) = (\psi_+, \psi_-)^{\mathrm{T}}$:

$$i\frac{\partial \mathbf{\Psi}}{\partial t} = -\frac{1}{2}\nabla^2 \mathbf{\Psi} - \sigma_1 \mathcal{R}(x,y)\mathbf{\Psi} + \sigma_3 \Omega \mathbf{\Psi} + g(\mathbf{\Psi}^\dagger \mathbf{\Psi})\mathbf{\Psi}, \quad (1)$$

where $\nabla^2 = \partial^2/\partial x^2 + \partial^2/\partial y^2$ is the Laplacian; $\sigma_1$ and $\sigma_3$ are the Pauli matrices; $\Omega$ accounts for amplitude of Zeeman splitting that introduces the asymmetry between two components; the parameter $g$ characterizes the strength of the intra- and inter-species interactions that for simplicity is assumed here equal and attractive (repulsive) for $g < 0$ ($g > 0$). We assume that no external potential is explicitly present in the system, but that nontrivial topology arises from spatial modulation of the Rabi coupling between $\psi_+$ and $\psi_-$ components described by the function $\mathcal{R}(x,y)$. As the simplest possible structures, we consider 1D or 2D SSH Rabi lattices, consisting of spots with locally increased coupling strength [see Fig. 1(a) and 5(a) for examples of such lattices], i.e. $\mathcal{R}(x,y) = \rho \sum_{m,n} e^{-[(x-x_m)^2 + (y-y_n)^2]/\sigma^2}$, with amplitude $\rho = 8$ and characteristic width $\sigma = 0.5$, where nodes $(x_m, y_n)$ of the lattice are arranged such as to form 1D line or 2D square SSH structures. Further we consider both zero and nonzero values of Zeeman splitting.

It should be stressed that by introducing the matrix $\mathcal{S}$:

$$\mathcal{S} = \frac{1}{2^{1/2}}\begin{pmatrix} 1 & -1 \\ 1 & 1 \end{pmatrix}, \quad \mathcal{S}^{-1} = \frac{1}{2^{1/2}}\begin{pmatrix} 1 & 1 \\ -1 & 1 \end{pmatrix} \quad (2)$$

that diagonalizes Pauli matrix $\sigma_1$, i.e. $\mathcal{S}\sigma_1\mathcal{S}^{-1} = -\sigma_3$, one can transform the Eq. (1) [by multiplying it with matrix $\mathcal{S}$ and introducing the wavefunction $\mathbf{\Phi} = \mathcal{S}\mathbf{\Psi} = 2^{-1/2}(\psi_+ - \psi_-, \psi_+ + \psi_-)^{\mathrm{T}}$] into the equation for this new wavefunction:

$$i\frac{\partial \mathbf{\Phi}}{\partial t} = -\frac{1}{2}\nabla^2 \mathbf{\Phi} + \sigma_3 \mathcal{R}(x,y)\mathbf{\Phi} + \sigma_1 \Omega \mathbf{\Phi} + g(\mathbf{\Phi}^\dagger \mathbf{\Phi})\mathbf{\Phi}, \quad (3)$$

where nonlinear term maintains its functional form because Manakov nonlinearity is considered here. This transformation shows that the system with inhomogeneous Rabi coupling between two components is formally analogous to the system with specific Zeeman SSH-like lattice (to the best of our knowledge topological solitons in lattices of this type were never explored previously) with uniform coupling between components whose strength is determined by the amplitude of Zeeman splitting $\Omega$ from the original Eq. (1).

On the other hand, by introducing the matrix $\mathcal{Q}$ with coordinate-dependent elements:

$$\mathcal{Q} = \frac{(\Omega^2 + \mathcal{R}^2)^{-1/2}}{2}\begin{pmatrix} \Omega + (\Omega^2 + \mathcal{R}^2)^{1/2} & -\mathcal{R} \\ (\Omega^2 + \mathcal{R}^2)^{1/2} - \Omega & \mathcal{R} \end{pmatrix}, \quad (4)$$

whose inverse is given by

$$\mathcal{Q}^{-1} = \mathcal{R}^{-1}\begin{pmatrix} \mathcal{R} & \mathcal{R} \\ \Omega - (\Omega^2 + \mathcal{R}^2)^{1/2} & \Omega + (\Omega^2 + \mathcal{R}^2)^{1/2} \end{pmatrix}, \quad (5)$$

that diagonalizes the entire matrix $-\sigma_1 \mathcal{R} + \sigma_3 \Omega$ from evolution Eq. (1), i.e. $\mathcal{Q}(-\sigma_1\mathcal{R} + \sigma_3\Omega)\mathcal{Q}^{-1} = (\Omega^2 + \mathcal{R}^2)^{1/2}\sigma_3$, after multiplication of the latter equation with coordinate-dependent matrix $\mathcal{Q}$ and introduction of new wavefunction $\mathbf{X} = \mathcal{Q}\mathbf{\Psi}$, one can rewrite Eq. (1) in the following alternative form:

$$i\frac{\partial \mathbf{X}}{\partial t} = -\frac{1}{2}\nabla^2 \mathbf{X} - \mathcal{Q}\nabla\mathcal{Q}^{-1}\nabla\mathbf{X} - \frac{1}{2}\mathcal{Q}\nabla^2\mathcal{Q}^{-1}\mathbf{X} + \\ (\Omega^2 + \mathcal{R}^2)^{1/2}\sigma_3\mathbf{X} + g\mathcal{Q}([\mathcal{Q}^{-1}\mathbf{X}]^\dagger\mathcal{Q}^{-1}\mathbf{X})\mathcal{Q}^{-1}\mathbf{X}. \quad (6)$$

Eq. (6) shows that the system can be also considered as a system with Zeeman SSH-like lattice, but with coordinate-dependent coupling defined by $\mathcal{Q}\nabla\mathcal{Q}^{-1}$ and $\mathcal{Q}\nabla^2\mathcal{Q}^{-1}$ matrices, whose elements are significant because they involve derivatives of the function $\mathcal{R}$ that contains strongly localized spots (these matrix elements are rather cumbersome and are not presented here). This highlights that the effect of inhomogeneous Rabi coupling in Eq. (1) is not equivalent to the presence of usual potential. Moreover, because matrix $\mathcal{Q}$ is not a unitary matrix, the latter system will be also characterized by the inhomogeneous cubic nonlinearity that will impact soliton solutions. Further we concentrate on topological properties of the system (1).

The unit cell of the 1D SSH Rabi lattice includes two spots (dimer), where coupling between BEC species is enhanced [Fig. 1(a)]. The width of the unit cell is further set as $d_x = 4$. To introduce nontrivial topology into this system, we shift the "sites" in each unit cell by the distance $s$ in the opposite directions from the locations corresponding to equal spacing $d_x/2$ between all sites of the lattice (middle panel). For $s < 0$ [top panel in Fig. 1(a)] two sites in each dimer approach each other, while for $s > 0$ [bottom panel in Fig. 1(a)] they are moving away from each other towards the borders of the unit cell. We consider 1D lattice with 11 unit cells.

We also consider 2D generalization of this system. The square unit cell of the 2D SSH Rabi lattice includes already four spots [see Fig. 5(a)] and it has dimensions $d_x \times d_y$. Further we consider square unit cell with $d_x = d_y = 4$ and assume that pairs of spots on the diagonals of the unit cell can be shifted (from the locations that correspond to the 2D lattice with identical spacing $d_x/2 = d_y/2$ between all sites) in the opposite directions along the diagonals (in this case the shift of lattice sites along the $x$ and $y$ axes is identical and is equal to $s$). For $s > 0$ [Fig. 5(b)] the sites shift towards the corners of the unit cell, while at $s < 0$ they shift towards the centre of the unit cell. In 2D case Rabi SSH lattice that we consider had $5 \times 5$ unit cells (i.e. $100$ spots where coupling between components is enhanced).

**Linear edge states and Zak phase for 1D Rabi SSH lattice**

Shift of sites in 1D Rabi SSH lattice has profound effect on its linear spectrum and may induce topological phase. To show this, it is instructive to calculate linear eigenmodes of the 1D lattice that can be found in the form $\Psi(x,y,t) = \mathbf{w}(x,y)e^{-i\mu t}$, where $\mathbf{w} = (w_+, w_-)^{\mathrm{T}}$ describes the profile of the eigenmode with chemical potential (eigenvalue) $\mu$, with $w_\pm$ being real functions. Such modes satisfy the linear equation

$$\mu \mathbf{w} = -\frac{1}{2}\nabla^2 \mathbf{w} - \sigma_1 \mathcal{R}(x,y)\mathbf{w} + \sigma_3 \Omega \mathbf{w}, \qquad (7)$$

obtained from Eq. (1) using the above substitution and setting $g = 0$. We solved it using plane-wave expansion method. Figure 1(b) and 1(c) show two representative dependencies of the eigenvalues of all lattice modes on the shift $s$ in the absence ($\Omega = 0$) and in the presence ($\Omega = 1$) of Zeeman splitting, respectively. One can see two bands in the spectrum (a consequence of the fact that there are two sites in the unit cell). Black dots correspond to extended bulk states. Red dots correspond to edge states emerging in the gap in topologically nontrivial regime at $s > 0$. At $s < 0$ the gap opens as well, but without edge states in it, since in this case the system remains nontopological. The appearance of topological edge states is associated with nonzero topological invariant of the system – winding number, related to Zak phase – that can be calculated for each band $n$ of periodic (non-truncated) lattice using the expression [97]:

$$C_n = \frac{i}{2\pi} \int_{\mathrm{BZ}} \langle \mathbf{u}_{k,n}(x,y) | \partial_k | \mathbf{u}_{k,n}(x,y) \rangle dk, \qquad (8)$$

where $\mathbf{u}_{k,n}(x,y) = \mathbf{u}_{k,n}(x+d_x,y)$ is the $x$-periodic part of the Bloch function $\Psi(x,y,t) = \mathbf{u}_{k,n}(x,y)e^{ikx-i\mu t}$ of Rabi lattice, $k$ is the Bloch momentum, $\mathbf{u}_{k,n} = (u_{kn+}, u_{kn-})^{\mathrm{T}}$, and the integration is carried over the first Brillouin zone (BZ). Typical dependencies $\mu(k)$ for two lowest bands are presented in Fig. 2(a),(b) for negative and positive shifts $s$ at $\Omega = 1$. When $s \leq 0$ the winding number is zero for both bands – consequently, no edge states appear in the spectrum of truncated Rabi lattice in Fig. 1(c) in this regime. In contrast, for $s > 0$ one gets nonzero winding numbers $C_1 = +1$ and $C_2 = -1$ for lowest two bands [see Fig. 2(b)], that indicates that edge states can emerge in this regime in the topological gap between these two bands. This picture holds for various values of Zeeman splitting $\Omega > 0$. Increasing $\Omega$ leads to downshift of the spectrum in Fig. 1(c) and to slight broadening of the bands, while their structure does not change qualitatively.

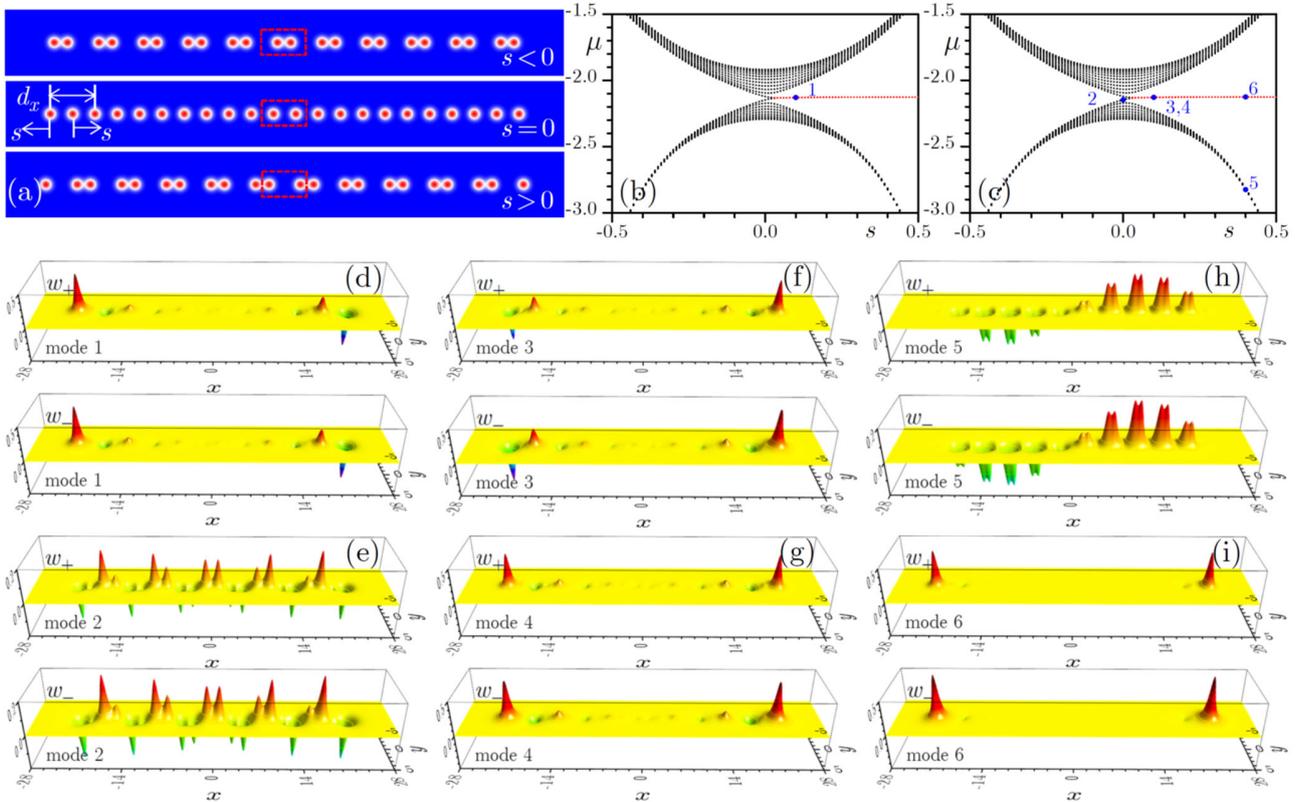

Fig. 1. (a) Schematic illustrations of truncated Rabi lattice $\mathcal{R}(x,y)$ with shifts $s=-0.4$ (top), $s=0$ (middle), and $s=+0.4$ (bottom). The red dashed rectangle indicates the lattice unit cell. Linear spectrum of truncated Rabi lattice versus shift $s$ for Zeeman splitting $\Omega=0$ (b) and $\Omega=1$ (c). Edge states are shown with red dots. Examples of profiles of linear eigenmodes of the Rabi lattice in topological (d), (f), (g), (i) and nontopological (e), (h) phases corresponding to the blue dots in (b) and (c). Both $w_+$ and $w_-$ components are shown.

Examples of linear modes supported by truncated Rabi lattice are presented in Fig. 1(d)-(i). These modes correspond to blue dots in Fig. 1(b) and 1(c) and they are enumerated accordingly. Here we show both $w_+$ and $w_-$ components that feature similar degree of localization. The modes are normalized such that $\iint \langle \mathbf{w}|\mathbf{w}\rangle dxdy = 1$. Because our lattice contains an integer number of the unit cells, the edge states emerge simultaneously at both edges of the lattice, leading to their out-of-phase [Fig. 1(d) and 1(f)] and in-phase [Fig. 1(g) and 1(i)] combinations. Since we consider large lattices, these combinations are practically degenerate and have nearly identical chemical potentials $\mu$ for not too small shifts $s$ of the lattice sites. The localization of topological edge states shown in Fig. 1(d),(f),(g),(i) increases with increase of $s$. All modes belonging to bulk bands [Fig. 1(e) and 1(h)] are delocalized. It should be mentioned that due to symmetry of Eq. (7), at $\Omega=0$ the $w_\pm$ components feature exactly the same shapes, while nonzero $\Omega$ introduces the asymmetry between them, c.f. Fig. 1(d) and 1(f). The latter can be quantified by the parameter

$$R_\pm = \frac{N_\pm}{N_+ + N_-} \qquad (9)$$

defined through norms of two components $N_\pm = \iint w_\pm^2 dxdy$ of solution. The dependencies $R_\pm(\Omega)$ are depicted in Fig. 2(c), and they show that for positive $\Omega$ the $w_-$ component becomes stronger than the $w_+$ one, while for negative $\Omega$ the situation is reversed. Increasing $|\Omega|$ leads to more and more pronounced asymmetry.

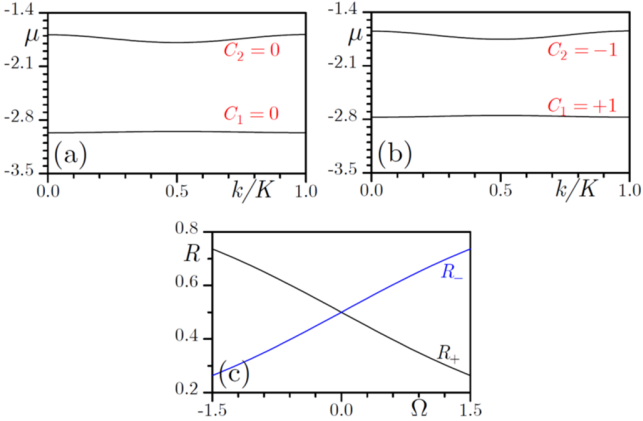

Fig. 2. Two lowest bands of the 1D Rabi lattice for $s=-0.4$ (a) and $s=+0.4$ (b), at Zeeman splitting $\Omega=1$. Winding numbers for each band are shown with red letters in each panel. (c) Asymmetry ratio $R_\pm$ for linear edge states, defined as per Eq. (9), versus $\Omega$ in topological regime at $s=0.4$.

**Edge solitons in 1D Rabi SSH lattice**

Having shown that Rabi SSH lattice can support localized linear edge states of topological origin, we now turn to their nonlinear generalizations – edge solitons that can bifurcate from linear edge states in topological gap under the action of interatomic interactions. Such solitons, $\mathbf{\Psi}(x,y,t) = \mathbf{w}(x,y)e^{-i\mu t}$, where $\mathbf{w} = (w_+, w_-)^T$, can be obtained from the nonlinear equation following from Eq. (1) using Newton method:

$$\mu \mathbf{w} = -\frac{1}{2}\nabla^2 \mathbf{w} - \sigma_1 \mathcal{R}(x,y)\mathbf{w} + \sigma_3 \Omega \mathbf{w} + g(\mathbf{w}^\dagger \mathbf{w})\mathbf{w}. \qquad (10)$$

Here we consider both repulsive and attractive interactions corresponding to $g=+1$ and $g=-1$, respectively. Edge soliton families characterized by the dependence of total norm $N = \iint \langle \mathbf{w}|\mathbf{w}\rangle dxdy$ on chemical potential $\mu$ are presented in Fig. 3(a) for representative case of $\Omega=1$, $s=0.4$, where red lines correspond to repulsive interactions, while black and blue lines correspond to attractive interactions. Grey areas in Fig. 3(a) correspond to allowed bands. One can see that edge solitons bifurcate from linear edge states, as their norm vanishes exactly in the point of bifurcation. The sign of interactions determines the direction of bifurcation. The norm of soliton increases away from the bifurcation point, until soliton enters into one of the allowed bands (depending on sign of $g$), where it couples with bulk modes and exhibits delocalization. Localization of solitons can be characterized by the form-factor $\chi$ (the quantity that is inversely proportional to soliton width) calculated using the formula

$$\chi^2 = N^{-2} \iint \langle \mathbf{w}|\mathbf{w}\rangle^2 dxdy. \qquad (11)$$

The larger is the form-factor, the better is the localization of soliton. Representative dependencies $\chi(\mu)$ are presented in Fig. 3(b). In the bifurcation point the form-factor of soliton coincides with that of the linear edge state. Close to the bifurcation point attractive interactions (black line) initially lead to enhancement of soliton localization, while repulsive interactions (red line) lead to slight soliton expansion. Form-factor notably decreases inside allowed bands, where soliton acquires long tails due to coupling with bulk modes.

Properties similar to properties of solitons in inhomogeneous Rabi coupling landscapes were encountered for 1D optical edge solitons in usual SSH lattices without Rabi coupling, e.g. in Refs. [62,63] illustrating bifurcation of solitons from linear edge states under the action of nonlinearity for the lattice consisting of dimers, and in Ref. [64] for the lattice consisting of trimers. All these works considered scalar systems with usual SSH-like optical potentials, in contrast to spinor system considered here. Nevertheless, 1D solitons obtained in [62-64] show qualitatively similar behaviour to solitons in atomic system with inhomogeneous Rabi coupling considered here – namely, their amplitude also increases away from the bifurcation point and they delocalize when their propagation constants shift into allowed bands of the system.

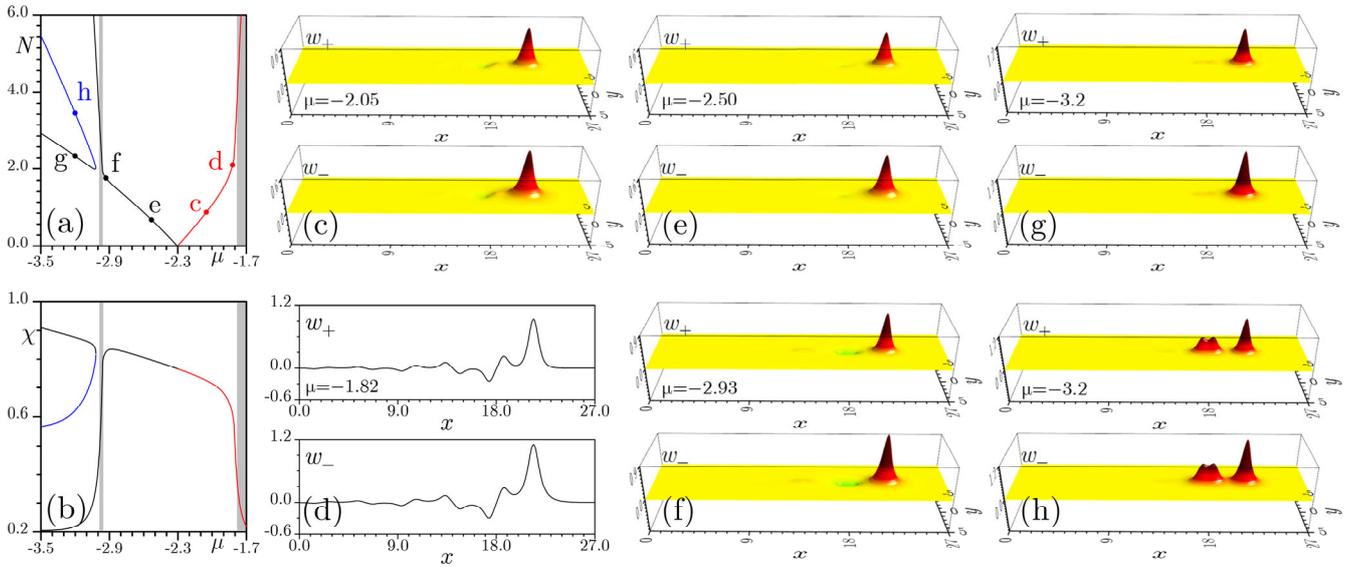

Fig. 3. Families of edge solitons in 1D Rabi SSH lattice and their profiles. Norm $N$ (a) and form factor $\chi$ (b) versus chemical potential $\mu$ for different families of solitons bifurcating from linear topological edge states, as well as solitons in semi-infinite gap. Red (black or blue) curves correspond to $g=+1$ ($g=-1$), grey regions show bulk bands. (c-h) Representative profiles of solitons corresponding to the coloured dots in (a), among them Fig. 3(d) displays the $x$ cross-section of the soliton to highlight the structure of soliton tails. All branches shown with solid lines correspond to stable states. In all cases $\Omega=1$, $s=0.4$.

In addition to edge solitons in topological gap [the region between two allowed grey bands in Fig. 3(a)], we also found edge solitons in the semi-infinite gap for attractive interactions. While topological solitons are thresholdless and exist even when their norm $N \to 0$, solitons in the semi-infinite gap exist above threshold norm, even though one of the families (the one with lower norm) looks like a continuation of the edge soliton family in semi-infinite gap. It should be stressed that solitons in topological and semi-infinite gaps differ in their internal structure. Representative soliton profiles are shown in Fig. 3(c)-3(h), whose labels correspond to labels near the dots in Fig. 3(a). Figures 3(c)-3(f) display solitons inside the topological gap, while Fig. 3(g) and 3(h) show solitons in the semi-infinite gap. We use continuous model in our simulations that takes into account all details of Rabi coupling landscape and complex interplay between nonlinearity and dispersion (i.e. we are far from tight-binding model assuming invariable modal shapes and considering only varying amplitudes on different "sites" of SSH-like structure). On this reason the $y$ profile of soliton changes in accordance with local value of amplitude [see an example of $x$ cross-section of profile for topological soliton denoted by point "d" in Fig. 3(d)]. For $\Omega=1$ the solitons have unequal $w_+$ and $w_-$ components, by analogy with properties of linear modes in this system. Gradual formation of long tails inside the lattice for $\mu$ values close to the edges of the topological gap is particularly well visible in Fig. 3(d). In solitons from topological gap, the field changes its sign in sites belonging to neighbouring dimers [Fig. 3(c)], but in solitons from semi-infinite gap the field does not change its sign in the above sites [Fig. 3(g) and 3(h)]. Notice also that two soliton families found in the semi-infinite gap merge near the gap edge.

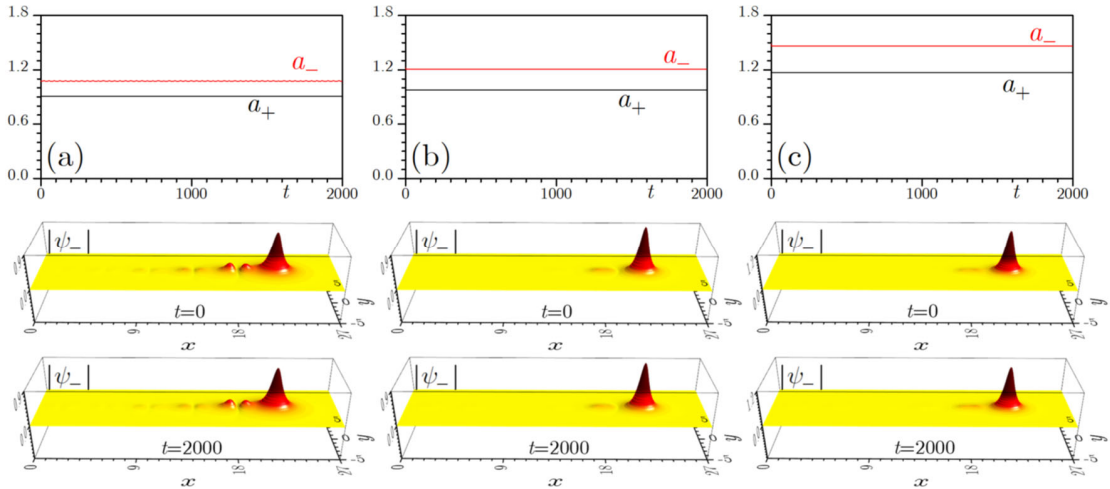

Fig. 4. Stable evolution of edge solitons from topological gap at $\mu=-1.85$ for repulsive interactions (a), at $\mu=-2.85$ for attractive interactions (b), and evolution of stable soliton from black branch in semi-infinite gap with $\mu=-3.1$ for attractive interactions (c). Only $|\psi_-|$ distributions

are shown. Corresponding maximal amplitudes of both components are shown in the top row as functions of time. In all cases $s=0.4$, $\Omega = 1$.

To confirm stability of solitons discussed above we modelled their evolution in the presence of various initial weak perturbations using Gross-Pitaevskii equation (1). The evolution of such perturbed states up to times $t \sim 10^4$ indicate that for our parameters all branches of solitons found in the gaps (topological or semi-infinite ones) are stable, including weakly localized states near topological gap edges. Examples of stable evolution of solitons belonging to the topological (semi-infinite) gap are shown in Fig. 4(a) and 4(b) (Fig. 4(c)). Only the initial (middle row) and final distributions (bottom row) are shown here for $|\psi_-|$ component, while top row shows evolution of maximal amplitude $a_\pm = \max|\psi_\pm|$ with time. As one can see, such states are robust, including weakly localized state in Fig. 4(a). Stability of such states is a noteworthy result, especially in attractive BEC, taking into account that it is achieved only due to modulation of coupling, without using usual external potentials.

**Linear corner states supported by 2D Rabi SSH lattice**

In this section we consider 2D generalization of the Rabi SSH lattice. As mentioned above, unit cells of such lattices contain four sites, whose shift along the diagonal of the unit cell allows to realize topologically nontrivial structure, see Fig. 5(a)-5(c) illustrating uniform lattice, topologically nontrivial structure obtained at $s>0$, and topologically trivial structure with $s<0$, respectively. 2D Rabi SSH lattice is a realization of higher-order topological insulator and it affords the possibility to study the properties of corner states supported by inhomogeneous coupling landscape $\mathcal{R}(x,y)$. Such corner states emerge in truncated 2D lattices for positive shifts $s>0$, when sites in the unit lattice cells are displaced towards their corners. The transformation of linear spectrum of 2D Rabi lattice as a function of $s$ is presented in Fig. 5(d). The lattice that we consider here contains $5 \times 5$ unit cells, each cell involves four domains with locally increased coupling. At $s>0$ one can observe the appearance of four bulk bands and one edge band in the spectrum. Strongly localized corner states depicted in Fig. 5(d) by red dots emerge only at sufficiently large shift $s>0.35$, in the third gap. No localized states were found in non-topological regime at $s<0$. Representative examples of linear states corresponding to the blue dots in Fig. 5(d) are presented in Fig. 5(e)-5(h). Among them, Fig. 5(e) illustrates an example of bulk mode at $s=0$, the edge mode is shown in Fig. 5(f) (the eigenvalues of such modes actually also form the band in the spectrum), while Fig. 5(g) and 5(h) display two different examples of topological corner states at $s=0.4$. Due to $\mathcal{C}_4$ discrete rotational symmetry of the lattice, there exist four practically degenerate corner states – here we show only two of them. Just as in the 1D case, nonzero Zeeman splitting $\Omega$ introduces asymmetry between $w_+$ and $w_-$ components.

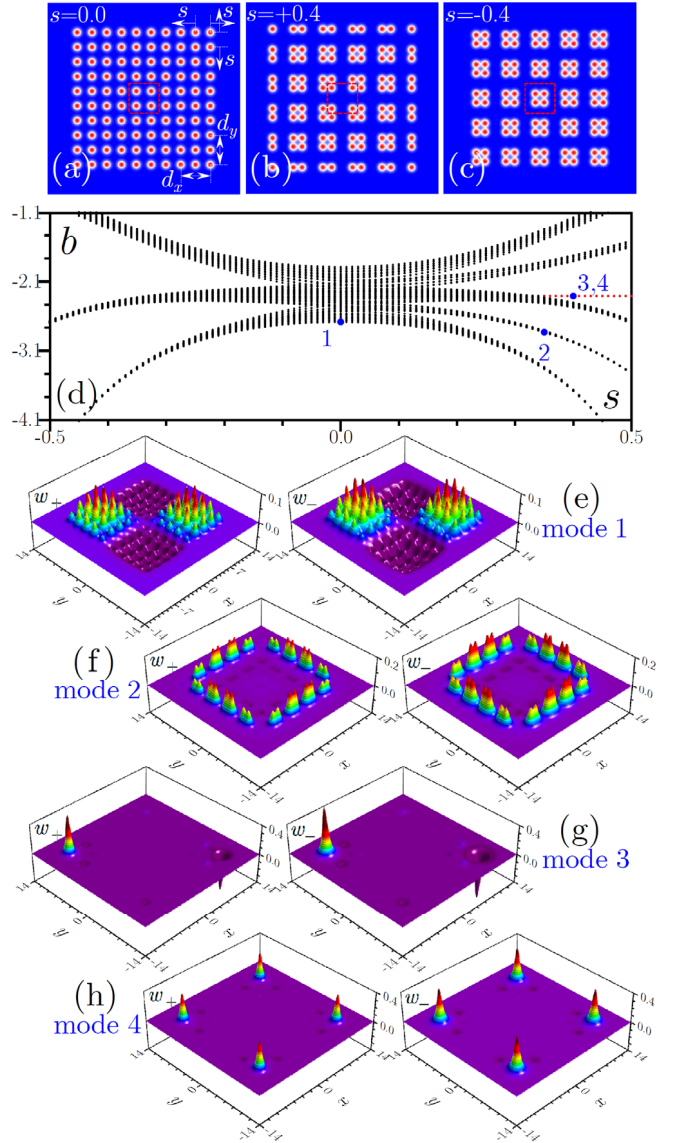

Fig. 5. Examples of 2D Rabi SSH lattices corresponding to shifts $s=0$ (a), $s=0.4$ (b), and $s=-0.4$ (c). The red dashed rectangle indicates the lattice unit cell. (d) Transformation of the spectrum of linear modes of 2D Rabi lattice with increasing shift of its sites $s$ for Zeeman splitting $\Omega = 1$. Eigenvalues of bulk and edge modes are shown black, while eigenvalues of topological corner modes are shown red. Representative extended modes of Rabi lattice (e),(f) corresponding to dots 1 and 2 in (d), and topological modes (g),(h) corresponding to dots 3 and 4 in (d).

Topological properties of the 2D Rabi SSH lattice are characterized by so-called 2D polarization, defined for a periodic lattice as [98]:

$$P_x = \frac{i}{S} \iint_{\text{BZ}} \langle \mathbf{u}_{\mathbf{k},n}(x,y) | \partial_{k_x} | \mathbf{u}_{\mathbf{k},n}(x,y) \rangle dk_x dk_y,$$
$$P_y = \frac{i}{S} \iint_{\text{BZ}} \langle \mathbf{u}_{\mathbf{k},n}(x,y) | \partial_{k_y} | \mathbf{u}_{\mathbf{k},n}(x,y) \rangle dk_x dk_y,$$
(12)

where $\mathbf{u}_{\mathbf{k},n}(x,y) = \mathbf{u}_{\mathbf{k},n}(x+d_x, y+d_y)$ is the periodic part of the 2D Bloch function $\Psi = \mathbf{u}_{\mathbf{k},n} e^{ik_x x + ik_y y - i\mu t}$ of the lattice, and $S$ is the area of the first Brillouin zone, $\mathbf{k} = (k_x, k_y)$ is the Bloch momentum. Topologically nontrivial phase is characterized by nonzero polarization $P_x = P_y = 1/2$ for topologically nontrivial gap where corner state appears, while in nontopological regime polarization is zero, for details see [98].

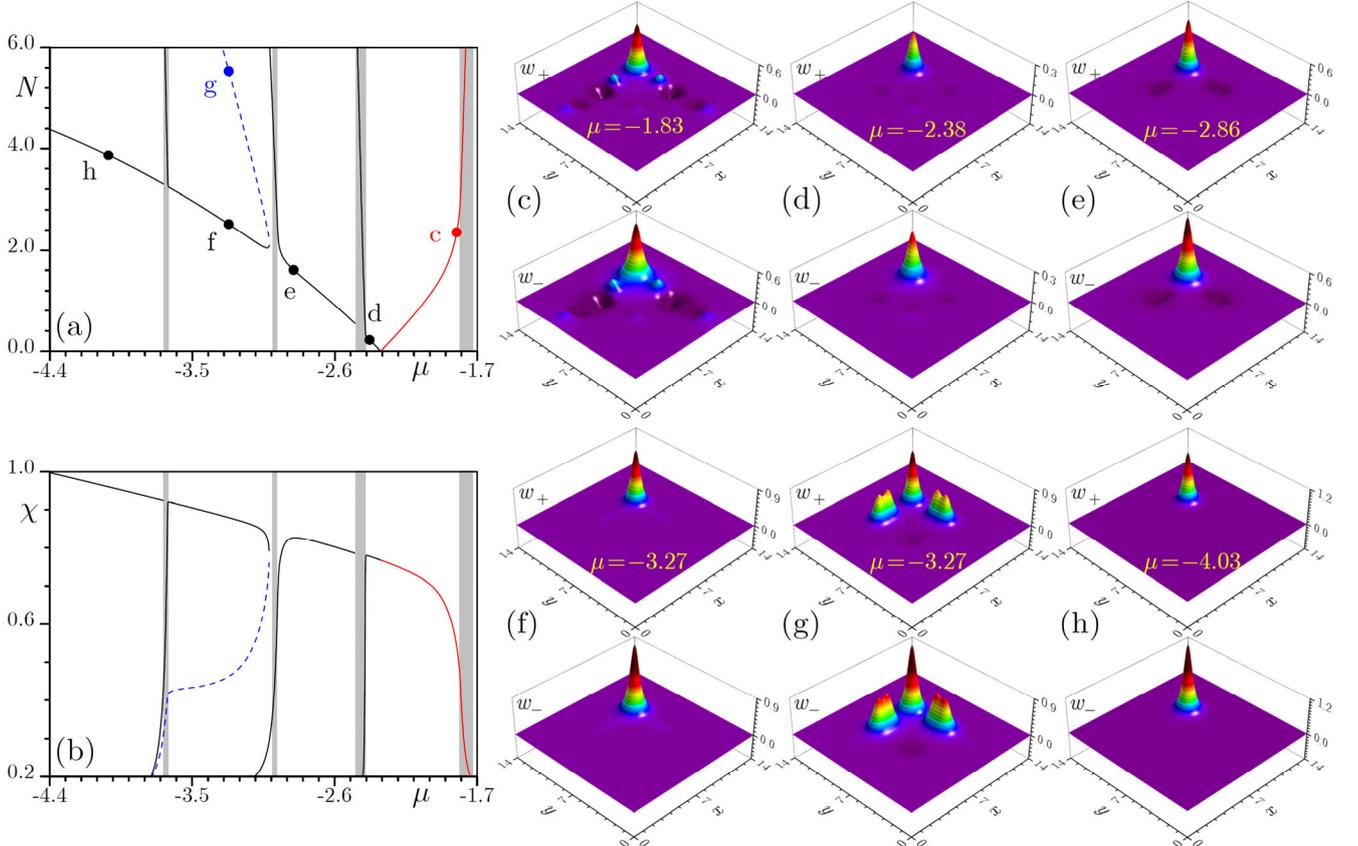

Fig. 6. Norm $N$ (a) and form factor $\chi$ (b) versus chemical potential $\mu$ for different families of corner solitons bifurcating from linear topological corner state for $\Omega = 1$ and $s = 0.4$. Red (black or blue) curves corresponding to $g = +1$ ($g = -1$), shaded regions show bulk and edge bands. Panels (c-h) display examples of solitons in different gaps of the spectrum corresponding to the dots in panel (a). Branches shown with solid lines are stable, while those shown with dashed lines are unstable.

**Topological corner solitons in 2D Rabi SSH lattice**

In this section we obtain the families of 2D topological corner solitons that can bifurcate from their linear counterparts at $g \neq 0$. Their profiles are described by Eq. (10) with $\mathcal{R}(x,y)$ function describing 2D Rabi lattice. Norm $N$ of corner states as a function of chemical potential $\mu$ is shown in Fig. 6(a) in different gaps of the spectrum. Corner solitons belonging to red (black or blue) branches are supported by repulsive (attractive) interatomic interactions. Corner solitons bifurcate from linear corner states in right outermost gap in Fig. 6(a). As in the 1D case, when corner soliton enters into the bulk or edge band, it strongly couples with corresponding states and becomes delocalized. This is accompanied by rapid growth of norm. The branches of corner solitons found for attractive nonlinearity in all gaps to the left of topological gap exist above threshold norm. Such branches may join with other branches near gap edges, see for example dashed blue curve joining with black one in Fig. 6(a). The dependence of soliton form-factor $\chi$ on chemical potential $\mu$ is shown in Fig. 6(b). One can observe overall growth of localization of corner soliton with decrease of $\mu$, interrupted by the regions of abrupt delocalization, when $\mu$ enters into the bands. For example, solitons deeply in semi-infinite gap [see example in Fig. 6(h)] are well localized, but they have different structure of tails from corner solitons in right topological gap. Examples of corner solitons from different gaps are shown in Fig. 6(c)-6(h). Among them, solitons from topological gap feature staggered tails, see Fig. 6(c) and 6(d) that are most pronounced in repulsive BEC. Solitons from black and blue branches in the second gap (from the left) in Fig. 6(a) differ in population of edge sites adjacent to the corner one, for example, in state shown in Fig. 6(g) this population is much stronger than for state in Fig. 6(f) for the same $\mu$. Notice that in all solitons shown here $w_+$ and $w_-$ components have different amplitudes due to nonzero Zeeman splitting.

Direct simulations of evolution of perturbed corner solitons show that the majority of obtained branches are stable in both attractive and repulsive BEC. This is interesting result once again confirming that spatially inhomogeneous Rabi coupling has strong stabilizing action, in particular for states of topological origin. Example of stable evolution of soliton with $\mu = -1.9$ supported in topological gap by repulsive interactions is shown in Fig. 7(a), while evolution of stable soliton from adjacent gap with $\mu = -2.7$ supported by attractive interactions is depicted in Fig. 7(b). In both cases, corner solitons survive over long time intervals and their amplitude shows only small oscillations due to initially imposed perturbation. The example of decay of soliton with $\mu = -3.56$ from blue dashed branch in attractive medium is shown in Fig. 7(c), where one can see irregular amplitude jumps and radiation into the bulk of the lattice.

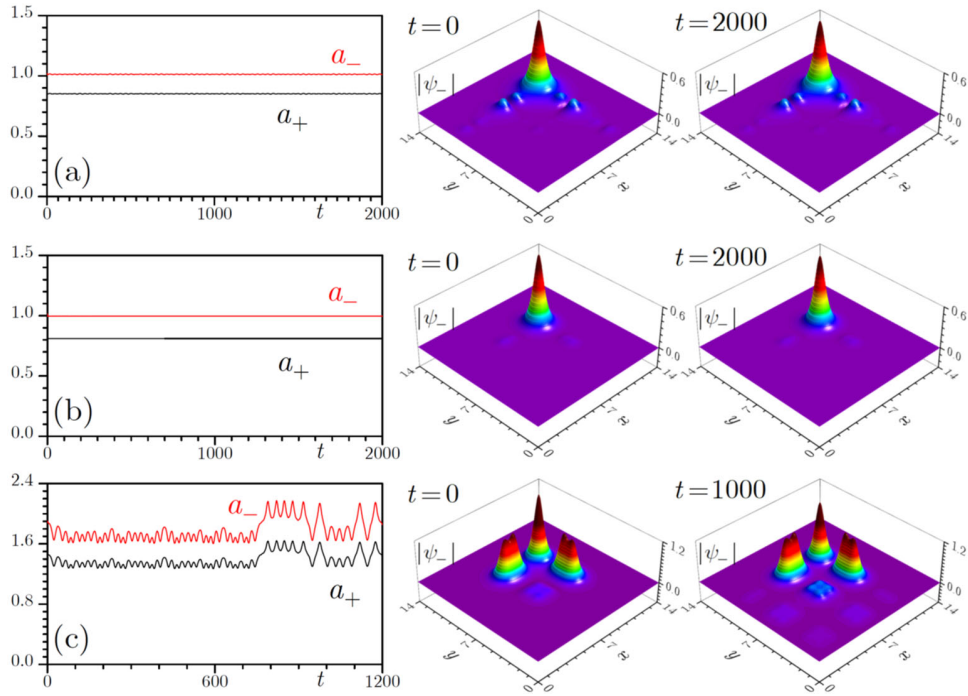

Fig. 7. Examples of stable evolution of corner solitons (a) with $\mu = -1.90$ in repulsive BEC and (b) with $\mu = -2.70$ in attractive BEC. (c) Example of unstable evolution of soliton with $\mu = -3.56$ belonging to blue branch in Fig. 6(a) in attractive case. Peak amplitudes $a_\pm = \max|\psi_\pm|$ are shown as functions of time in the left column, while right panels compare initial and final profiles of $\psi_-$ component.

## Conclusions

Summarizing we proposed a system, where nontrivial topology can be created by spatial modulation of Rabi coupling strength between two BEC species. If such modulation forms 1D or 2D SSH lattice, the latter can support localized edge or corner modes, whose existence is guaranteed by nontrivial topology of the bands of such Rabi lattice, despite the fact that we do not explicitly use external linear potential. In the presence of attractive or repulsive interatomic interactions, such edge and corner states give rise to exceptionally robust edge and corner solitons, whose stability is a result of spatial modulation of the coupling strength. Our study paves the way for realization of new topological phases in atomic systems.

## Acknowledgements

The authors are very grateful to Prof. V. V. Konotop for fruitful discussion of this work. This work was funded by the grant of National Natural Science Foundation of China (NSFC) (11805145) and China Scholarship Council (CSC) (202006965016). Y.V.K acknowledges the research project FFUU-2021-0003 of the Institute of Spectroscopy of the Russian Academy of Sciences.